\documentclass[prd,twocolumn,aps,floats]{revtex4}

\usepackage[dvips]{graphicx}
\usepackage{amssymb}
\usepackage{amsmath}
\begin{document}

\title{Tracking $f(R)$ cosmology}

\author{Mahmood Roshan and Fatimah Shojai}

\affiliation{Department of Physics, University of Tehran, Tehran,
Iran }
\begin{abstract}
Metric $f(R)$ gravity theories are conformally equivalent to
models of quintessence in which matter is coupled to dark energy.
We derive a condition for stable tracker solution for metric
$f(R)$ gravity in the Einstein frame. We find that tracker
solutions with $-0.361<\omega_{\varphi}<1$ exist if
$0<\Gamma<0.217$ and $\frac{d}{dt} \ \ln f'(\tilde{R})>0$, where
$\Gamma=\frac{V_{\varphi\varphi}V}{V_{\varphi}^{2}}$ is
dimensionless function, $\omega_{\varphi}$ is the equation of
state parameter of the scalar field and $\tilde{R}$ refers to
Jordan frame's curvature scalar. Also, we show that there exists
$f(\tilde{R})$ gravity models which have tracking behavior in the
Einstein frame and so the curvature of space time is decreasing
with time while they lead to the solutions in the Jordan frame
that the curvature of space time can be increasing with time.
\end{abstract}
\maketitle

\section{Introduction}

   Fourth order $f(R)$ gravity theories can be considered
 as a candidate for solving the major challenge of
cosmology i.e. the late time accelerated expansion of the universe
\cite{sahni}, without introducing any exotic matter sources
\cite{faraoni}(and references therein). These theories have
particular features among the other modified theories of gravity.
$f(R)$ modifications to GR appear in the low-energy effective
actions of quantum gravity and the quantization of fields in
curved spacetime.  These theories suggest a completely geometric
origin for both the early time inflation and the late time cosmic
acceleration.

These theories are conformally related to GR with a
self-interacting scalar field \cite{barrow}. Although these models
can rise to a natural acceleration mechanism, there exists some
features in them which make their viability dubitable. For
examples these models predict an amount $\frac{1}{2}$ for PPN
parameter $\gamma$, which is a gross violation of the experimental
bound $|\gamma-1|<2.3 \ 10^{-5}$ \cite{bertotti}. Albeit chameleon
$f(R)$ gravity models can pass the solar system tests but in the
sense of cosmological considerations these theories are
observationally indistinguishable from a cosmological constant
\cite{amendola}.

Any way, our purpose here is to find out a condition for stable
tracker solutions for metric $f(R)$ gravity models in the Einstein
frame. Although we pass from fourth order gravity to scalar-tensor
gravity in which equations are mathematically simpler, but its
physical relevance is still controversial \cite{capozziello}.
However, following standard procedures one should not conclude
equivalently about the physical relation between the results. It
has been demonstrated that passing from one frame to another can
alert the physical meanings of the results (see
\cite{capozzielloplb} and references therein). For example the
stability of solutions can be completely different in the two
frames \cite{gorini}. However, we pay our attention just to the
Einstein frame.

Consider the general action of these theories in Jordan frame
\begin{eqnarray}
S_{J}=\int d^{4}x
\sqrt{-\tilde{g}}\left[\frac{f(\tilde{R})}{12\alpha^{2}}+\mathcal{L}_{m}(\tilde{g}_{\mu\nu})\right].
\end{eqnarray}
Where $\alpha=\sqrt{\frac{4\pi G}{3}}$ and all tilded quantities
are in Jordan frame. Under the conformal transformation
$g_{\mu\nu}=e^{2\alpha \varphi}\tilde{g}_{\mu\nu}$
 , where $\varphi=\frac{1}{2\alpha}\ln f'$ and prime denotes derivative with respect to $\tilde{R}$ , we obtain the Einstein frame
action
\begin{eqnarray}
\begin{split}
S_{E}=\int d^{4}x
\sqrt{-g}[\frac{R}{12\alpha^{2}}-&\frac{1}{2}(\nabla\varphi)^{2}
\\ &-V(\varphi)+\mathcal{L}_{m}(e^{-2\alpha\varphi}g_{\mu\nu})].
\end{split}
\end{eqnarray}
Where $V=(\tilde{R}f'-f)/12\alpha^{2}f'^{2}$. We see that in the
Einstein frame the scalar field couples minimally to gravity but
couples conformally to matter fields via the function
$e^{-2\alpha\varphi}$. For a spatially flat FRW space-time the
modified Friedmann equations and the equation of motion of scalar
field are given by
\begin{eqnarray}\label{friedmann}
H^{2}=2\alpha^{2}(\rho_{\varphi}+\rho_{m})
\end{eqnarray}
\begin{eqnarray}\label{friedmann1}
\dot{H}=-3\alpha^{2}\left[\rho_{m}(1+\omega_{m})+\rho_{\varphi}(1+\omega_{\varphi})\right]
\end{eqnarray}
\begin{eqnarray}\label{scalar field}
\ddot{\varphi}+3H\dot{\varphi}+V_{\varphi}=\alpha(1-3\omega_{m})\rho_{m}
\end{eqnarray}
Where $\rho_{m}$ and $p_{m}$ are the energy density and pressure
of cosmic fluid in the Einstein frame. Also,
$\rho_{\varphi}=\frac{1}{2}\dot{\varphi}^{2}+V(\varphi)$ and
$p_{\varphi}=\frac{1}{2}\dot{\varphi}^{2}-V(\varphi)$ represent
the energy density and pressure of the scalar field.

 In Einstein
frame, the scalar field and the cosmic fluid satisfy the
conservation equations
\begin{eqnarray}\label{rhophidot}
\dot{\rho}_{\varphi}+3H(1+\omega_{\varphi})\rho_{\varphi}=\alpha\dot{\varphi}(\rho_{m}-3p_{m})
\end{eqnarray}
\begin{eqnarray}\label{rhomdot}
\dot{\rho}_{m}+3H(1+\omega_{m})\rho_{m}=-\alpha\dot{\varphi}(\rho_{m}-3p_{m})
\end{eqnarray}
and the energy density of matter $\rho_{m}$, pressure $p_{m}$,
cosmic time $t$ and scale factor $a$ are related to their Jordan
frame counterparts through \cite{faraonibook}
\begin{eqnarray}
\rho_{m}=e^{-4\alpha\varphi}\tilde{\rho}_{m}  \ \
p_{m}=e^{-4\alpha\varphi}\tilde{p}_{m} \ \
dt=e^{\alpha\varphi}d\tilde{t} \ \ a=e^{\alpha\varphi}\tilde{a}
\end{eqnarray}

It is clear from equation (\ref{rhophidot}) that the evolution of
scalar field is not determined only by its potential energy since
there is a coupling to matter. Furthermore, such couplings give
rise to additional forces on matter particles in addition to
gravity. During the matter dominated phase, by using equation
(\ref{rhophidot}), one can introduce an effective potential as
follows
\begin{eqnarray}\label{Veffective}
V_{eff}(\varphi)=V(\varphi)+\rho^{*}e^{-\alpha\varphi}
\end{eqnarray}
Where $\rho^{*}$ is a conserved quantity in the Einstein frame
\cite{khoury}, which is related to $\rho_{m}$ via the relation
$\rho_{m}=\rho^{*}e^{-\alpha\varphi}$.

\section{tracking solutions}

In this section we find the condition for having tracking
solutions. We confine our attention to the case
$\dot{\Omega}_{\varphi}>0$ which is satisfactory from the
astrophysical point of view \cite{johri} and the generalization to
the case $\dot{\Omega}_{\varphi}<0$ can be done with similar
considerations. In the uncoupled quintessence model this condition
requires that $\omega_{\varphi}<\omega_{m}$ and $\omega_{\varphi}$
be nearly constant. But here we have
\begin{eqnarray}\label{omegadot}
\dot{\Omega}_{\varphi}=3H(\omega_{m}-\omega_{\varphi})\Omega_{m}\Omega_{\varphi}+
\alpha(1-3\omega_{m})\dot{\varphi}\Omega_{m}.
\end{eqnarray}
where
\begin{eqnarray}\label{phidot2}
 \dot{\varphi}^{2}=(1+\omega_{\varphi})\rho_{\varphi}=(1+\omega_{\varphi})\frac{H^{2}}{2\alpha^{2}}\Omega_{\varphi},
\end{eqnarray}
Using this equation to eliminate $\dot{\varphi}$, equation
\eqref{omegadot} becomes
\begin{eqnarray}
\begin{split}
\dot{\Omega}_{\varphi}=H\sqrt{\frac{1+\omega_{\varphi}}{2}}&\Omega_{m}\Omega_{\varphi}^{1/2}
(\pm(1-3\omega_{m})\\&
+3(\omega_{m}-\omega_{\varphi})\sqrt{\frac{2\Omega_{\varphi}}{1+\omega_{\varphi}}}).
\end{split}
\end{eqnarray}

At tracking era $\Omega_{\varphi}\ll\Omega_{m}$, so we can ignore
the second term in the above equation. In the matter dominated
era, this shows that the condition of $\dot{\Omega}_{\varphi}>0$
means that $\dot{\varphi}>0$ and so we select the plus sign. Also
it is not necessary that $\omega_{\varphi}$ be negative and it can
takes any value between -1 and 1(provided that $\Omega_{\varphi}$
is very small such that $\Omega_{\varphi}<0.1$). To obtain the
tracker equation we can express the equation of motion of scalar
field into the following form
\begin{eqnarray}\label{V1}
\frac{V_{\varphi}}{V}=-3\alpha\sqrt{\frac{2(1+\omega_{\varphi})}{\Omega_{\varphi}}}
\left[1+\frac{1}{6}\frac{x'}{x}\right]
+\frac{2\alpha}{1-\omega_{\varphi}}\frac{\Omega_{m}}{\Omega_{\varphi}},
\end{eqnarray}
where $x=\frac{1+\omega_{\varphi}}{1-\omega_{\varphi}}$ is the
ratio of the kinematic to the potential energy for $\varphi$ and
prime denotes derivative with respect to $\ln a$. Therefore, for a
tracker solution ($\omega_{\varphi}\simeq const$) the tracker
condition becomes
\begin{eqnarray}\label{V2}
\frac{V_{\varphi}}{V}\simeq-3\alpha\sqrt{\frac{2(1+\omega_{\varphi})}{\Omega_{\varphi}}}
+\frac{2\alpha}{1-\omega_{\varphi}}\frac{\Omega_{m}}{\Omega_{\varphi}},
\end{eqnarray}
By taking the time derivative of this equation for tracker
solution, we get
\begin{eqnarray}\label{V3}
\frac{d}{dt}(\frac{V_{\varphi}}{V})\simeq\frac{3\alpha\sqrt{1+\omega_{\varphi}}\dot{\Omega}_{\varphi}}{\Omega_{\varphi}^{3/2}}
(1-\frac{\sqrt{8}\Omega_{\varphi}^{-1/2}}{3(1-\omega_{\varphi})\sqrt{1+\omega_{\varphi}}}).
\end{eqnarray}
Therefore, during tracking, the second term is dominated and we
have $\frac{d}{dt}(\frac{V_{\varphi}}{V})<0$ or equivalently
$\Gamma<1$, where $\Gamma$ is a dimensionless function defined as
$\Gamma=\frac{V_{\varphi\varphi}V}{V'^{2}}$ \cite{steinhardt} and
is nearly constant at tracker period. This condition is completely
different from the tracker condition in the uncoupled quintessence
model ($\Gamma>1$) \cite{steinhardt}. By taking the time
derivative of equation (\ref{V1}) and combining with Friedmann
equations and equation (\ref{V1}) itself, we obtain the following
relation
\begin{eqnarray}\label{Gamma1}
\begin{split}
\Gamma=&1-\frac{2}{1+\omega_{\varphi}}\frac{\tilde{\tilde{y}}}{(6+\tilde{y})^{2}}-
\frac{1-\omega_{\varphi}}{2(1+\omega_{\varphi})}\frac{\tilde{y}}{6+\tilde{y}}\\
&+\frac{3(\omega_{m}-\omega_{\varphi})}{1+\omega_{\varphi}}\frac{1-\Omega_{\varphi}}{6+\tilde{y}}
+\frac{\lambda\Omega_{m}\Omega_{\varphi}^{-1/2}}{2(6+\tilde{y})} \\
&+\frac{(\omega_{\varphi}-1)\lambda\Omega_{m}\Omega_{\varphi}^{-1/2}}{2(1+\omega_{\varphi})}.
\end{split}
\end{eqnarray}
Where $\tilde{x}=\frac{d\ln x}{d\ln a}$,
$\tilde{y}=\tilde{x}-\lambda\Omega_{m}\Omega_{\varphi}^{-1/2}$,
$\tilde{\tilde{y}}=\frac{d\tilde{y}}{d\ln a}$ and
$\lambda=(1-3\omega_{m})\sqrt{\frac{2}{1+\omega_{\varphi}}}$. From
now we take $\omega_{m}=0$ because we are considering the matter
dominated era where metric $f(R)$ gravity behaves like an
interacting quintessence. Since $\Omega_{\varphi}\ll\Omega_{m}$
during tracking, a useful relation can be obtained for $\Gamma$ by
expanding it as a power series of $\Omega_{\varphi}$ i.e.
\begin{eqnarray}\label{Gamma2}
\begin{split}
\Gamma\simeq &
\frac{1}{2}-\sqrt{\frac{1+\omega_{\varphi}}{8}}+\left[\Sigma_{1}+\Sigma_{2}
\tilde{x}\right]\Omega_{\varphi}^{1/2}\\
&+\left[\Sigma_{3}+\Sigma_{4}\tilde{x}+\Sigma_{5}\tilde{x}^{2}-\tilde{\tilde{x}}\right]\Omega_{\varphi}+
O(\Omega_{\varphi}^{3/2}).
\end{split}
\end{eqnarray}
Where
\begin{eqnarray}\label{A}
\begin{split}
&\Sigma_{1}=6\omega_{\varphi}/\sqrt{2(1+\omega_{\varphi})}-3\sqrt{(1+\omega_{\varphi})/2}-3(1+\omega_{\varphi}),\\
&\Sigma_{2}=(1-\omega_{\varphi})\sqrt{8(1+\omega_{\varphi})}-\sqrt{(1+\omega_{\varphi})/8}-\frac{1+\omega_{\varphi}}{2},\\
&\Sigma_{3}=18\omega_{\varphi}-9-(28+27\omega_{\varphi})\frac{1+\omega_{\varphi}}{2},\\
&\Sigma_{4}=-\frac{\omega_{\varphi}}{2}-\frac{3}{4}\frac{(1+\omega_{\varphi})^{3/2}}{\sqrt{2}},\\
&\Sigma_{5}=-\frac{3}{2}-\frac{9(1+\omega_{\varphi})^{3/2}}{\sqrt{2}},
\end{split}
\end{eqnarray}

Therefore for the tracker solution (assuming $\Gamma$ is nearly
constant and $\Omega_{\varphi}\ll\Omega_{m}$) to the first order
in $\omega_{\varphi}$, we have
\begin{eqnarray}\label{Gamma3}
\Gamma\simeq \frac{1}{2}-\sqrt{\frac{1+\omega_{\varphi}}{8}}.
\end{eqnarray}
 This equation shows that for interacting quintessence corresponding to the Einstein frame of $f(R)$ gravity, tracking
occurs if $\Gamma<\frac{1}{2}$, although there exists another
constraints which limit this condition.

 Another constraint which limits this
interval, comes from the stability requirement. We require that
for any solution for which the equation of state parameter of dark
energy is different from tracking parameter, $\omega_{0}$, by a
small amount such as $\delta\omega$, then $\delta\omega$ decays
with time and the solution joins to the tracker solution.

\section{Stability of the tracking solutions}
Now we want to check the stability of tracking solutions with
constant $\omega_{\varphi}$. Consider a solution which is
perturbed form the tracker solution,
$\omega_{\varphi}=\omega_{0}$, by an amount $\delta\omega$. Then
we expand equation (\ref{Gamma2}) to lowest order in
$\delta\omega$. It should be noted that $\delta\omega$ and
$\Omega_{\varphi}$ have the same order of magnitude and so one can
neglect terms such as $\Omega_{\varphi}\delta''\omega$ or
$\Omega_{\varphi}^{3/2}\delta\omega$. However, we do not neglect
the terms containing powers smaller than one in $\Omega_{\varphi}$
and we apply the limit $\Omega_{\varphi}\rightarrow 0$ to final
solution for $\delta\omega$. By using the equations for
$\Sigma_{i}$ and expanding them to the first order of
$\delta\omega$ we have
$\Sigma_{i}=\Sigma_{i}(\omega_{0})+\sigma_{i}\delta\omega$, where
\begin{eqnarray}
\begin{split}
&\sigma_{1}=\frac{6}{\sqrt{2(1+\omega_{0})}}-
\frac{3\omega_{0}}{\sqrt{2}(1+\omega_{0})^{3/2}}-\frac{3}{\sqrt{8(1+\omega_{0})}}-3,\\
&\sigma_{2}=
\frac{\omega_{0}-1}{\sqrt{32}(1+\omega_{0})^{3/2}}+\frac{1}{\sqrt{32(1+\omega_{0})}},\\
&\sigma_{3}=-\frac{27}{\sqrt{8}}(1+\omega_{0}),\\
&\sigma_{4}=27\sqrt{\frac{1+\omega_{0}}{2}}+\frac{28+27\omega_{0}}{\sqrt{8(1+\omega_{0})}}.
\end{split}
\end{eqnarray}

Finally, by using these equations we obtain
\begin{eqnarray}\label{deltaomega}
\delta''\omega+\eta \delta'\omega+\xi \delta\omega=0.
\end{eqnarray}
where
\begin{eqnarray}
\begin{split}
&\eta=-[\Sigma_{2}(\omega_{0})\Omega_{\varphi}^{-1/2}+\Sigma_{3}(\omega_{0})],\\
&\xi=\frac{\omega_{0}^{2}-1}{2}[\Sigma_{1}(\omega_{0})\Omega_{\varphi}^{-1/2}
-\frac{\Omega_{\varphi}^{-1}}{\sqrt{32(1+\omega_{0})}}],
\end{split}
\end{eqnarray}
the solution of the equation (\ref{deltaomega}) is
\begin{eqnarray}
\delta\omega\sim a^{-\frac{\eta\pm i\sqrt{4\xi-\eta^{2}}}{2}}.
\end{eqnarray}
In order to have a decaying $\delta\omega$, $\eta$ should be
positive even if the quantity under square root is negative. One
can easily show that $\eta$ is positive , for any $\omega_{0}$ in
the interval $-1<\omega_{0}<1$, if $\Omega_{\varphi}>0.025$. When
$\Omega_{\varphi}<0.025$ then the interval of $\omega_{0}$ for
which $\eta$ is positive becomes tighter. We select the smallest
interval corresponding to $\Omega_{\varphi}\sim0$. It is easy to
show that this interval is
\begin{eqnarray}\label{tracking parameter}
-0.361<\omega_{0}<1,
\end{eqnarray}
 By taking into account equation
(\ref{Gamma3}) we obtain the final condition for having tracker
solutions
\begin{eqnarray}\label{Gamma5}
0<\Gamma<0.217.
\end{eqnarray}
In a similar way, the conditions such as \eqref{Gamma5} has been
derived for the existence of tracker solutions for K-essence in
\cite{chiba} and also for quintessence and k-essence in a general
cosmological background in \cite{das}.

The condition \eqref{Gamma5} should not to be confused with the
condition $\Gamma>1$ which has been appeared in the chameleon
scalar tensor theory \cite{khoury}. $f(R)$ gravity models can be
considered as a chameleon theory for which $\beta$ is negative and
is equal to $-1/\sqrt{6}$ \cite{cemberanos}. In order to find
tracker condition, the sign of $\beta$ is important. Since in the
chameleon scalar tensor theory, this coupling constant is
positive, the second term in the equation \eqref{V2} is negative.
By a similar calculation one can easily verify that the tracking
condition is $\Gamma>1$. Note that the condition $\dot{\varphi}>0$
is again required for having increasing density parameter of dark
energy.

\section{properties of tracking solutions}
 In the uncoupled quintessence model $\rho_{\varphi}\sim a^{-3(1+\omega_{\varphi})}$ and always
$\dot{\rho}_{m}<0$ and $\dot{\rho}_{\varphi}<0$. But here the
conservation equation of the dark energy density, equation
\eqref{rhophidot}, is not integrable even if equation of state
parameter is nearly constant. So the tracking behavior is not
clear. Furthermore, it is clear from (\ref{rhophidot}) that if
$\omega_{\varphi}$ is nearly constant then $\dot{\rho}_{\varphi}$
can be positive or negative. Also it is not obvious that whether
$|\dot{\rho}_{\varphi}|<|\dot{\rho}_{m}|$ during tracking, like
that of uncoupled quintessence model, or not.

Here, we want to clarify these ambiguities and explore the
tracking period with some more details. To do this, we can write
equation (\ref{rhophidot}) as follows
\begin{eqnarray}\label{rhophidot2}
\dot{\rho}_{\varphi}=\alpha
\rho_{\varphi}\dot{\varphi}\left(\frac{\Omega_{_{m}}}{\Omega_{\varphi}}
-3\sqrt{\frac{2(1+\omega_{\varphi})}{\Omega_{\varphi}}}\right).
\end{eqnarray}
Therefore, at the beginning of the matter dominated era the first
term is dominated and so dark energy density is increasing. On the
other hand, when $\Omega_{\varphi}$ becomes larger then the second
term is dominated and so $\rho_{\varphi}$ is decreasing. Thus,
there exist a maximum at the time evolution of $\rho_{\varphi}$.
We require that the dark energy density has not significant role
in the matter dominated era and its role is important for us at
late times. So, it is necessary to show that at this maximum,
$\rho_{\varphi}$ is smaller enough than the matter density. Let us
assume that the dark energy density takes its maximum at the time
$t^{*}$, then by using equation (\ref{rhophidot2}) we obtain
\begin{eqnarray}
\Omega_{m}|_{t^{*}}=\frac{-\beta+\sqrt{\beta^{2}+4\beta}}{2},
\end{eqnarray}
where $\beta=18(1+\omega_{\varphi})$. Now, by using the condition
(\ref{tracking parameter}) we have
\begin{eqnarray}
0.925<\Omega_{m}|_{t^{*}}<0.974.
\end{eqnarray}
Thus, the dark energy density at $t^{*}$ is small compared to the
matter density.

Moreover, it can be shown that $\rho_{\varphi}$ decreases at a
slower rate than $\rho_{m}$, so $t^{*}$ is smaller than the time
at which the matter and dark energy densities are equal. Firstly,
let us to show that in this era
$|\dot{\rho}_{\varphi}|<|\dot{\rho}_{m}|$.

After $t^{*}$, by using equations \eqref{rhophidot} and
\eqref{rhomdot}, the condition
$|\dot{\rho}_{\varphi}|<|\dot{\rho}_{m}|$ can be written as
\begin{eqnarray}\label{decrease}
(1+\omega_{\varphi})\frac{\Omega_{\varphi}}{1-\Omega_{\varphi}}-\frac{\sqrt{2(1+\omega_{\varphi})}}{3}\Omega_{\varphi}^{1/2}<1.
\end{eqnarray}
During tracking the density parameter of dark energy is small and
it can be shown that the condition \eqref{decrease} is satisfied
for any $-1<\omega_{\varphi}<1$. After tracking, in order to have
$|\dot{\rho}_{\varphi}|<|\dot{\rho}_{m}|$, as in the uncoupled
quintessence, it is necessary that $\omega_{\varphi}$ be
decreasing with time. By using equations \eqref{Veffective} and
\eqref{V1}, it can be shown that $\dot{\omega_{\varphi}}<0$ if
\begin{equation}\label{negativeomega}
\frac{1}{V}\left|\frac{dV_{eff}}{d\varphi}\right|<3\alpha\sqrt{\frac{2(1+\omega_{\varphi})}{\Omega_{\varphi}}},
\end{equation}
after tracking $\Omega_{\varphi}$ is not negligible compared to 1
so $\sqrt{2(1+\omega_{\varphi})/\Omega_{\varphi}}\sim O(1)$.
Consequently, equation \eqref{negativeomega} can be written as
\begin{equation}
M_{p\l}\frac{1}{V}\left|\frac{dV_{eff}}{d\varphi}\right|<1.
\end{equation}
in which $M_{pl}=(\sqrt{6}\alpha)^{-1}$.

 Now, we discuss the convergence to the tracker solution for
which $\omega_{\varphi}$ is nearly constant, in more detail. It is
required that the evolution of the tracking dark energy density be
insensitive to the initial conditions and any perturbation from
tracking dark energy density should be decreasing with time. This
is the main goal of introducing the tracking solutions
\cite{steinhardt}.

\begin{figure}[!t]
\hspace{0pt}\rotatebox{0}{\resizebox{.4\textwidth}{!}{\includegraphics{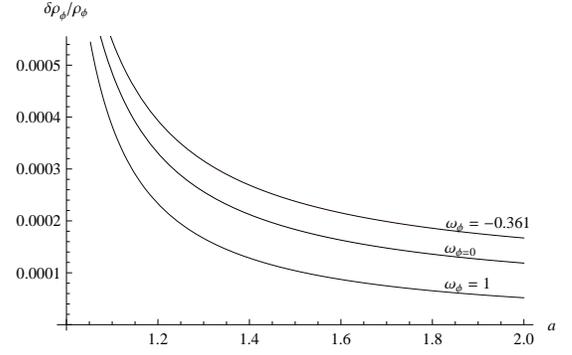}}}
\vspace{0pt}{\caption{The cosmic scale factor at matter-radiation
equality is normalized to 1. Assume that $\chi(1)\sim 10^{-3}$ and
$\Omega_{\varphi}(1)\sim 10^{-3}$.}\label{deltarho}}
\end{figure}

Suppose that $\varphi$ is perturbed from the tracking solution by
an small amount $\delta\varphi$. Any perturbation in the
"position" and "velocity" of the scalar field $\varphi$ produces a
perturbation, $\delta\rho_{\varphi}$, in the dark energy density
$\rho_{\varphi}$. In order to show that the fractional
perturbation $\frac{\delta\rho_{\varphi}}{\rho_{\varphi}}$ decays
with time, we write the continuity equation of the dark energy
density \eqref{rhophidot} as follows

\begin{eqnarray}
\rho'_{\varphi}+3(1+\omega_{\varphi})\rho_{\varphi}=\sqrt{\frac{(1+\omega_{\varphi})}{2}}\Omega_{\varphi}^{1/2}\rho_{m},
\end{eqnarray}
so a small perturbation $\delta\rho_{\varphi}$ will satisfy
\begin{eqnarray}
\begin{split}
\delta\rho'_{\varphi}+3&(1+\omega_{\varphi})\delta\rho_{\varphi}=\\&\sqrt{\frac{(1+\omega_{\varphi})}{8}}
\left(\Omega_{\varphi}^{-1/2}-\Omega_{\varphi}^{1/2}\right)
\delta\rho_{\varphi},
\end{split}
\end{eqnarray}
Now by using the above equations, one can write a differential
equation for $\chi=\frac{\delta\rho_{\varphi}}{\rho_{\varphi}}$ as
follows
\begin{eqnarray}\label{chi}
\chi'+\sqrt{\frac{(1+\omega_{\varphi})}{8}}
\left(\Omega_{\varphi}^{-1/2}-\Omega_{\varphi}^{1/2}\right)\chi=0.
\end{eqnarray}
 In order to find out $\Omega_{\varphi}$ during the tracking era, let us
write equation \eqref{omegadot} in the form
\begin{eqnarray}
\begin{split}
\Omega'_{\varphi}+3\omega_{\varphi}&\left(\Omega_{\varphi}-\Omega_{\varphi}^{2}\right)+\\&\sqrt{\frac{(1+\omega_{\varphi})}{2}}
\left(\Omega_{\varphi}^{3/2}-\Omega_{\varphi}^{1/2}\right)=0,
\end{split}
\end{eqnarray}
Taking into account that the equation of state parameter of dark
energy is nearly constant for the tracker solutions and also
$\Omega_{\varphi}\ll\Omega_{m}$ in the tracking era, this equation
can be solved analytically when the term containing
$\Omega_{\varphi}^{2}$ is ignored. For the sake of simplicity in
solving equation \eqref{chi}, here we will ignore also the term
containing $\Omega_{\varphi}^{3/2}$. With this approximation, the
above equation has the following solution
\begin{eqnarray}
\sqrt{\Omega_{\varphi}}=\frac{\sqrt{2(1+\omega_{\varphi})}}{6\omega_{\varphi}}+\gamma_{0}a^{-\frac{3\omega_{\varphi}}{2}},
\end{eqnarray}
Where $\gamma_{0}$ is an integration constant. By substituting
this solution in the equation \eqref{chi}, the solution for
$\chi(a)$ is

\begin{eqnarray}
\chi(a)=\frac{\delta\rho_{\varphi}}{\rho_{\varphi}}=\frac{\chi_{0}}{\sqrt{\Omega_{\varphi}}}y^{\frac{18\omega_{\varphi}^{2}-\omega_{\varphi}-1}{18\omega_{\varphi}^{2}}}
e^{-\frac{\gamma_{0}\sqrt{2(1+\omega_{\varphi})}}{6\omega_{\varphi}}y}.
\end{eqnarray}
in which $\chi_{0}$ is an integration constant and
$y=a^{-3\omega_{\varphi}/2}$. Also the corresponding solutions for
$\omega_{\varphi}=0$ are
\begin{eqnarray}
\begin{split}
&\sqrt{\Omega_{\varphi}}=\frac{\ln a}{\sqrt{8}}+\gamma_{0},\\
&\chi(a)=\frac{\delta\rho_{\varphi}}{\rho_{\varphi}}=\frac{\chi_{0}}{\sqrt{\Omega_{\varphi}}}\
 a^{\frac{\gamma_{0}}{\sqrt{8}}} \ e^{(\ln a^{1/4})^{2}}.
 \end{split}
\end{eqnarray}
This fractional perturbation has been plotted for various
$\omega_{\varphi}$, in Fig.\ref{deltarho}. We see that
$\frac{\delta\rho_{\varphi}}{\rho_{\varphi}}$ is a decreasing
function with time. As mentioned before, in the uncoupled
quintessence model $\frac{\delta\rho_{\varphi}}{\rho_{\varphi}}$
is constant during the tracking period, but here this ratio
quickly reaches to zero. Therefore, the evolution of the dark
energy density is insensitive to the initial conditions for the
dark energy density.
\section{DISCUSSION}
In this paper we have derived the conditions for fourth order
$f(R)$ gravity models to have tracker solutions in the Einstein
frame. The tracker solutions exist if $\omega_{\varphi}$ is nearly
constant and also i) $0<\Gamma<0.217$ and ii)
$\dot{\varphi}=\frac{d}{dt} \ \ln f'(\tilde{R})>0$.

For exploring the condition ii) further, let us consider only
positive-definite forms of $f'(\tilde{R})$, because the conformal
transformation $g_{\mu\nu}=e^{2\alpha
\varphi}\tilde{g}_{\mu\nu}$is singular for $f'(\tilde{R})=0$. This
forms of $f(R)$ also are necessary in order to have positive
effective gravitational coupling in the Jordan frame
\cite{sotiriou}. Also, suppose that $f''(\tilde{R})>0$ which is
required for Ricci scalar stability \cite{dolgovvafaraoni}. Ricci
scalar instability may appear if the matter-energy density (or
equivalently the scalar curvature) is large enough compared with
the matter density of the universe, for more details see
\cite{capozzielloGRG}. By these assumptions, the condition ii)
reduces to
\begin{eqnarray}
\frac{d\tilde{R}}{d\tilde{t}}>0
\end{eqnarray}

Therefore, although the scalar curvature in the Einstein frame is
decreasing and $f(R)$ gravity model can show tracking behavior,
the scalar curvature  of the Jordan frame can be increasing.

\begin{figure}
\begin{center}
\includegraphics[scale=0.8]{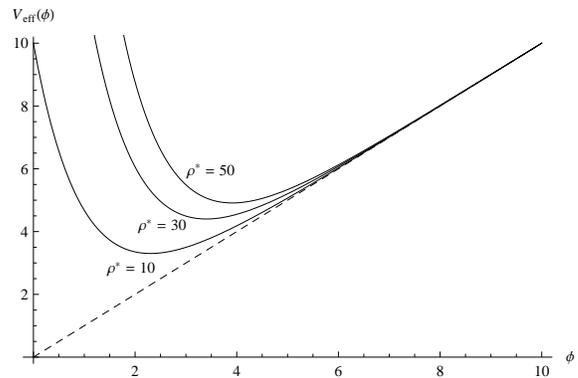}
%\hspace{0pt}\rotatebox{0}{\resizebox{.4\textwidth}{!}{\includegraphics{fig1}}}
%\vspace{0pt}{
\caption{Effective potential of the model
\eqref{f(r)2},
$V_{eff}(\varphi)=V_{0}\varphi^{n}+\rho^{*}e^{-\alpha\varphi}$, for
several values of $\rho^{*}$. Units and constants have been
suppressed i.e. $V_{0}=n=\alpha=1$(note that for $n=1$ the tracker
condition is satisfied).}
\end{center}
\label{veffect}
\end{figure}

Now, as an example, we want to find out a class of $f(\tilde{R})$
models which have constant $\Gamma$ and so can show tracker
behavior in the Einstein frame. Only for two classes of potentials
the function $\Gamma$ is constant, power law and exponential
potentials. Exponential potentials lead to $\Gamma=1$, which
cannot pass the tracking condition. Thus, we consider the power
law potentials. In this case, $V(\varphi)$ takes the form
\begin{eqnarray}\label{vgammaconst}
V(\varphi)=V_{0}\varphi^{\frac{1}{1-\Gamma}}=V_{0}\varphi^{n}
\end{eqnarray}
Where $V_{0}$ is a positive integration constant. Taking into
account the condition i) then $1<n<1.28$. Now, by using
$V=(\tilde{R}f'-f)/12\alpha^{2}f'^{2}$ and equation
\eqref{vgammaconst} we get
\begin{eqnarray}\label{f(R)}
\tilde{R}f'(\tilde{R})-f(\tilde{R})+U_{0}f'(\tilde{R})^{2}[\ln
f'(\tilde{R})]^{n}=0
\end{eqnarray}
Where $U_{0}=-12\alpha^{2}V_{0}/(2\alpha)^{n}$. Let us to solve
this equation for the case $\alpha\varphi\ll 1$, or equivalently
$\ln f'\ll1$. By supposing that
$f(\tilde{R})=\tilde{R}+\Psi(\tilde{R})$ and neglecting the terms
containing $\Psi(\tilde{R})$ with powers larger than one (note
that $n$ is of order unity), the equation \eqref{f(R)} reduces to
\begin{eqnarray}
\tilde{R}\Psi'(\tilde{R})-\Psi(\tilde{R})+U_{0}\Psi'(\tilde{R})^{n}=0
\end{eqnarray}
The solution of this differential equation is
$\mu^{2}(m-1)\left(\frac{\tilde{R}}{\mu^{2}}\right)^{m}$ and so
\begin{eqnarray}\label{f(r)2}
f(\tilde{R})=\tilde{R}-\mu^{2}(1-m)\left(\frac{\tilde{R}}{\mu^{2}}\right)^{m}
\end{eqnarray}
Where $m=\frac{n}{n-1}$ and
\begin{eqnarray}
\mu^{2}=\frac{-U_{0}}{[n^{m}(1-m)^{2}]^{\frac{1}{1-m}}}
\end{eqnarray}
Taking into account the condition i) one gets $m>4.60$. Now we
show that the condition ii) is satisfied for this model too. When
$\dot{\varphi}>0$ then we expect that the slop of the effective
potential which the scalar field experiences it is negative i.e.
$\frac{d}{d\varphi}V_{eff}<0$. The effective potential
corresponding to the model \eqref{f(r)2} is illustrated in
Fig.\ref{veffect}. By using equation \eqref{Veffective}, the
condition $\frac{d}{d\varphi}V_{eff}<0$ reduces to
\begin{eqnarray}
V_{\varphi}<\alpha\rho_{m}\leq\alpha\rho_{m}(t_{e})
\end{eqnarray}

In this model, there exists a free parameter $\mu$ which can be
sufficiently small to pass this condition and so, all condition
for having tracking solutions are satisfied.
It is important to note that the model \eqref{f(r)2} is similar to
the chameleon $f(\tilde{R})$ model which has been considered
before in the literature \cite{amendola}. But, in the case of the
chameleon $f(\tilde{R})$ gravity models, theories of the kind
\eqref{f(r)2} are compatible with observation in the range of the
parameter $0<m<0.25$ \cite{cemberanos}. Therefore, this class of
chameleon $f(\tilde{R})$ gravity models cannot pass the sufficient
condition for having tracker solutions.

 \section{acknowledgments} This work is partly
supported by a grant from university of Tehran and partly by a
grant from center of excellence of department of physics on the
structure of matter.


\newpage
   \begin{thebibliography}{99}
   \bibitem{sahni}
   V. Sahni and A. A. Staroboinsky, Int. J. Mod. Phys. D \textbf{9},
   373 (2000); T. Padmanabhan, Phys. Rept. \textbf{380},235 (2003).


   \bibitem{faraoni}
   P. Sotiriou and V. Faraoni ,arXiv:0805.1726.

   \bibitem{barrow}
   J. D. Barrow and S. Cotsakis, Phys. Lett. B 214, 515 (1988); K. I.
   Maeda, Phys. Rev. D \textbf{39}, 3159 (1989).

   \bibitem{bertotti}
   Bertotti, B., L. Iess, and P. Tortora, 2003, Nature \textbf{425},
   374.

   \bibitem{amendola}
   Amendola, L., R. Gannouji, D. Polarski, and S. Tsujikawa, 2007,
   Phys. Rev. D\textbf{75}, 083504. Amendola, L., D. Polarski, and S.
   Tsujikawa, 2007, Phys. Rev. Lett. \textbf{98}, 131302. Amendola,
   L., D. Polarski, and S. Tsujikawa, 2007, Int. J. Mod. Phys.
   D\textbf{16}, 1555.

   \bibitem{capozziello}
   Y.M. Cho, Class. Quantum Grav. \textbf{14} (1997) 2963, S.
   Capozziello, R. de Ritis, A.A. Marino, Class. Quantum Grav.
   \textbf{14} (1997)3243.

   \bibitem{capozzielloplb}
   S. Capozziello, S. Nojiri, S. D. Odintsov, A. Troisi, Phys. Lett.
   B \textbf{639} (2006) 135143
   \bibitem{gorini}
   V. Gorini, A. Kamenshchik, U. Moschella, V. Pasquier, A.
   Starobinsky, Phys. Rev. D \textbf{72} (2005) 103518.
   \bibitem{faraonibook}
  V. Faraoni, \textit{Cosmology in scalar-tensor gravity} (Kluwer Academic Publishers,
  Dordrecht,2004)
    \bibitem{khoury}
    Khoury, J., and A.Weltman, 2004a, Phys. Rev. D\textbf{69}, 044026.
    Khoury, J., and A. Weltman, 2004b, Phys. Rev. Lett. \textbf{93},
    171104.

   \bibitem{johri}
   P. G. Ferreira and M. Joyce, Phys. Rev. Lett. \textbf{79},4740
   (1997); L. Wang, R. R. Caldwell, J. P. Ostriker and P. Steinhardt,
   Astrophys. J. \textbf{530}, 17 (2000)

      \bibitem{steinhardt}
P. J. Steinhardt, L. Wang, and Ivaylo Zlatev, Phys.Rev.Lett.
\textbf{82}, 896-899 (1999), astro-ph/9807002; P. J. Steinhardt, L.
Wang, I. Zlatev, Phys.Rev. D \textbf{59}, 123504 (1999),
astro-ph/9812313.
      \bibitem{chiba}
   T. Chiba, Phys. Rev. D \textbf{66} (2002) 063514
 \bibitem{das}
   Rupam Das, Thomas W. Kephart, Robert J. Scherrer , Phys. Rev. D
   \textbf{74}:103515,2006

   \bibitem{cemberanos}
    Faulkner, T., M. Tegmark, E. F. Bunn, and Y. Mao, 2007, Phys. Rev.
   Cembranos, J. A. R., 2006, Phys. Rev. D\textbf{73}, 064029.
   D\textbf{76}, 063505. Starobinsky, A. A., 2007, JETP Lett.
   \textbf{86}, 157.
   \bibitem{sotiriou}
   T. P. Sotiriou and V. Faraoni, eprint gr-qc/0805.1726
   \bibitem{dolgovvafaraoni}
    Dolgov, A. D., and M. Kawasaki, 2003a, Phys.
   Lett. B573, 1. Faraoni, V., 2006a, Phys. Rev. D74, 104017.
   \bibitem{capozzielloGRG}
   S. Capozziello, M. Laurentis, S. Nojiri and S. D. Odintsove,
   eprint hep-th/808.1335

   \end{thebibliography}
\end{document}